\documentclass[conference]{IEEEtran}
\IEEEoverridecommandlockouts

\usepackage{cite}
\usepackage{tabularx}
\usepackage{amsmath,amssymb,amsfonts}
\usepackage[table,xcdraw]{xcolor}
\usepackage{algorithmic}
\usepackage{graphicx}
\usepackage{textcomp}
\usepackage{cleveref} % for \cref command
\usepackage{url}
\usepackage{adjustbox}

\usepackage{amsmath}
\usepackage{algorithm}
\usepackage{algorithmic}
\usepackage{float} % for the [H] option
\usepackage{xcolor}
\usepackage{array}
\begin{document}
%%%%%%%%%%%%%%%%%%%%%%%%%%%%%%%%%%%%%%%%%%%%%%%%%%%%%%%%%%%%%%%%%%%%%%%%%%%%%%%%%%%%%%
\title{
%Addressing Inaccuracy in Blind Power Identification Method through Optimized Initialization

%Cluster-BPI: Clustering Optimized Initialization for Accurate Blind Power Identification 

% Clustering Initialization for Accurate Blind Power Identification 
Fine-Grained Clustering-Based Power Identification for Multicores

%\vspace{-5mm}
\thanks{The authors would like to thank the following funding agencies: NSF grants 2219679
and 2219680.}
 }

\author{
    \IEEEauthorblockN{Mohamed R. Elshamy\IEEEauthorrefmark{1}, Mehdi Elahi\IEEEauthorrefmark{2}, Ahmad Patooghy\IEEEauthorrefmark{2}, and Abdel-Hameed A. Badawy\IEEEauthorrefmark{1}}
    \IEEEauthorblockA{\IEEEauthorrefmark{1}Klipsch School of ECE, New Mexico State University, Las Cruces, NM 88003, United States}
    \IEEEauthorblockA{\IEEEauthorrefmark{2}Computer Systems Technology, North Carolina A\&T State University, Greensboro, NC, United States}
    \IEEEauthorblockA{\IEEEauthorrefmark{1}\{elshamy, badawy\}@nmsu.edu, \IEEEauthorrefmark{2}melahi@aggies.ncat.edu, apatooghy@ncat.edu}
}

%%%%%%%%%%%%%%%%%%%%%%%%%%%%%%%%%%%%%%%%%%%%%%%%%%%%%%%%%%%%%%%%%%%%%%%%%%%%%%%%%%%%%%
\maketitle

\begin{abstract} 
Fine-grained power estimation in multicore Systems on Chips (SoCs) is crucial for efficient thermal management. BPI (Blind Power Identification) is a recent approach that determines the power consumption of different cores and the thermal model of the chip using only thermal sensor measurements and total power consumption. BPI relies on steady-state thermal data along with a naive initialization in its Non-negative Matrix Factorization (NMF) process, which negatively impacts the power estimation accuracy of BPI. This paper proposes a two-fold approach to reduce these impacts on BPI. First, this paper introduces an innovative approach for NMF initializing, \textit{i.e.}, density-oriented spatial clustering to identify centroid data points of active cores as initial values. This enhances BPI accuracy by focusing on dense regions in the dataset and excluding outlier data points. Second, it proposes the utilization of steady-state temperature data points to enhance the power estimation accuracy by leveraging the physical relationship between temperature and power consumption. Our extensive simulations of real-world cases demonstrate that our approach enhances BPI accuracy in estimating the power per core with no performance cost. For instance, in a four-core processor, the proposed approach reduces the error rate by $76\%$ compared to BPI and by $24\%$ compared to the state of the art in the literature, namely,  Blind Power Identification Steady State (BPISS). The results underline the potential of integrating advanced clustering techniques in thermal model identification, paving the way for more accurate and reliable thermal management in multicores and SoCs. 
\end{abstract}

%%%%%%%%%%%%%%%%%%%%%%%%%%%%%%%%%%%%%%%%%%%%%%%%%%%%%%%%%%%%%%%%%%%%%%%%%%%%%%%%%%%%%%
%%%%%%%%%%%%%%%%%%%%%%%%%%%%%%%%%%%%%%%%%%%%%%%%%%%%%%%%%%%%%%%%%%%%%%%%%%%%%%%%%%%%%%

\section{Introduction}
As the pace of Moore’s law decelerates, designers have sought alternative design methodologies, which led to the development of heterogeneous multicore architectures and the incorporation of specialized hardware units into a single chip, also known as Systems on Chip (SoCs)~\cite{b1, JIANG2024207}. However, this advancement has introduced challenges in thermal management, power consumption, and energy efficiency~\cite{b1}. 
 The increased power density coupled with constrained cooling options in SoCs has led to significant performance bottlenecks~\cite{b3, RANGARAJAN2023185}. 
 
Accurately estimating the power consumption of each core in a multicore processor is critical for effective thermal management and performance optimization. Precise power estimation improves the efficiency of dynamic voltage and frequency scaling (DVFS) and thermal throttling, maintaining safe temperature limits and enhancing processor reliability and lifespan. For instance, the Running Average Power Limit (RAPL) interface allows applications to measure power consumption~\cite{5599016}. However, these measurements are generally coarse-grained, providing only the overall power consumption of all cores, uncore units, and total package power. Inaccurate estimations can cause thermal hotspots, suboptimal performance, reduced lifespan, and increased cooling costs due to the need for additional cooling mechanisms~\cite{b4,b5,pr11123450}. Therefore, it is crucial to develop techniques and tools that enable fine-grained profiling of existing SoCs and the software they operate. This step is foundational for implementing effective power and thermal management strategies and designing next generation SoCs.

Blind Power Identification (BPI) techniques are used for managing power consumption and thermal behavior in multicore SoCs in cases where detailed pre-silicon models are not available~\cite{b4,b5}. These methods use statistical or machine learning approaches to infer power usage and heat generation from observable data like total power consumption and thermal sensor readings. However, BPI performance is highly dependent on the model initial conditions, leading to significant accuracy and robustness variability~\cite{b4,b5,b6}. This sensitivity poses challenges in environments where precise thermal management is vital, such as SoCs used in mobile applications. The variability step from Non-negative Matrix Factorization (NMF)~\cite{b13,10.1093/comjnl/bxab103} used to decompose aggregated sensor data into distinct thermal resistance and power matrices, essential for managing thermal outputs in multicore processors. However, NMF heavily relies on initial values, with traditional initialization methods often being too sensitive, resulting in inconsistent results and poor model accuracy.

% \textcolor{red}{since you have space, you can add a paragraph on the negative consequences on inaccurate power/thermal estimations, make sure you add references supporting claims.}

This paper presents an innovative approach for initializing NMF, crucial for BPI in multicore SoCs. By employing density-based spatial clustering of applications with noise (known as the DBSCAN clustering technique)~\cite{b7,10.5555/3001460.3001507}, our approach identifies cluster centroids from dense regions in the initial thermal data to be used as starting points NMF. This approach minimizes the impact of outliers and ensures NMF is initialized with optimal points from the main dataset, thereby enhancing robustness. Furthermore, leveraging steady-state temperatures for initialization improves the accuracy of power estimation by reflecting the true thermal behavior of the system. Our extensive simulations demonstrate that this approach significantly improves the precision of estimated power per core while maintaining computational efficiency.

The rest of this paper is organized as follows. Sections~\ref{Sec:related-work} and~\ref{sec:background} review related work and provide relevant background on thermal and power modeling. Section~\ref{Sec:Proposed-Init} introduces our proposed approach. Section~\ref{sec:results}  discusses the results obtained from our simulations, comparing the accuracy and efficiency of our method against existing techniques. Finally, Section~\ref{sec:conl} concludes the paper.
\section{Related Work}
\label{Sec:related-work}
% \subsection{Maintaining the Integrity of the Specifications}

Numerous studies have explored different techniques for modeling SoCs thermal and power aspects~\cite{b4,b5,b6,b8,b9}. These approaches typically employ a standard approach, aiming to discern the state space model connecting temperature and power according to Eqn.~\ref{Eq:1}.

\begin{equation}
    T_r(k) = AT_r(k-1) + BP(k).
    \label{Eq:1}
\end{equation}

In this model, \( T_r(k) \) and \( P(k) \) represent vectors denoting the temperature and power levels of SoC units at time \( k \), respectively, the matrices \( A \) and \( B \) encapsulate the physical relationship between power and temperature. While matrix \( A \) is the thermal conductance matrix that illustrates the system's natural response in the absence of power input, matrix \( B \) describes the system's forced response as a function of thermal capacitance and conductance. Both matrices \( A \) and \( B \) are square, and their dimensions correspond to the number of power sources.

The power sources align with the hardware units for which thermal measurements are available. All the previous studies~\cite{b4,b5,b6,b8,b9} aim to determine the power consumption of these units. Developing state space modeling matrices allows for accurate estimation and prediction of power consumption, mirroring the precision of the available thermal measurements. 

The state space model presented in Eq.~\ref{Eq:1} originates from the heat diffusion equation~\cite{b10}. This equation models the interaction between power and thermal properties by incorporating factors such as thermal conductivity, material density, and specific heat capacity. The formulation of the model in Eqn.~\ref{Eq:1} involves first performing a spatial discretization of the heat diffusion equation, followed by temporal discretization~\cite{b4,b5}. 

The difficulty in obtaining detailed power measurements has resulted in only a limited number of studies that offer valuable information about the power consumption and efficiency of various SoC hardware units under the strain of different software applications~\cite{b4,b5,b6,b8,b9}.

In this paper, we focus on a critical algorithm, BPI~\cite{b4,b5,b6}, which identifies thermal models and fine-grain power consumption of a chip using only data from thermal sensors and overall power consumption measurements. The accuracy of the algorithm estimations is aimed to be improved while maintaining the runtime. Reda~\textit{et al.}~\cite{b4} proposed the first version of BPI, which utilized the fast Independent Component Analysis (ICA) algorithm~\cite{761722} to initialize NMF by providing an initial guess of the factors that enhance convergence and performance by ensuring better separation of the underlying sources. Reda~\textit{et al.}~\cite{b5} proposed an enhanced BPI version with a different NMF initialization, utilizing the identity matrix for the thermal resistance matrix, as both matrices have their maximum values along the main diagonal. Said~\textit{et al.}~\cite{b6} introduced BPISS, an enhancement that initializes the thermal resistance matrix with the average steady state temperatures: non-diagonal elements for stressed cores and diagonal elements for cores with similar thermal characteristics. The core steady state temperature is used if no similar cores are found. The power matrix is initialized by dividing the total power based on each core temperature ratio by the sum of all core temperatures. This method better reflects the system physical characteristics and reduces error compared to identity matrix initialization.

% There are three general approaches to identifying the state space model:
%%%%%%%%%%%%%%%%%%%%%%%%%%%%%%%%%%%%%%%%%%%%%%%%%%%%%%%%%%%%%%%%%%%%%%%%%%%%%%%%%%%%%%
\section{Background}
\label{sec:background}
\subsection{Blind Power Identification Method}
The goal of the blind estimation problem is to determine the matrices \(A\) and \(B\) for a specific SoC, along with the power profiles \(P(k)\) in Eq.~\ref{Eq:1}. The BPI algorithm operates in two distinct phases, the initial phase is offline learning, where the system parameters \(A\), \(B\), and \(R\) are estimated using steady state measurements, $T_s$, and the total power at each interval, $P_{T_{s}}(k)$. This foundational phase sets the groundwork for the subsequent phase. The second phase is online learning that occurs during runtime using the runtime dataset, $T_r(k)$, and $P_{T_{r}}(k)$. At this stage, each core power consumption is estimated dynamically, enabling real-time adjustments and optimizations. The offline training step is required to be performed just once for each SoC to establish the modeling matrices.

% \begin{figure}%[htbp]
% \centering
% \includegraphics[width=1.03\linewidth]{Images/Fig1.png}
% \caption{Interactions and dataflow of the Blind Power Identification Algorithm}
% \label{fig:BPI_Phases}
% \end{figure}

To compute the \(B\) matrix in the first phase, determining the \(R\) matrix (the thermal resistance matrix) in the steady-state scenario is required~\cite{b5}. In this scenario, the steady-state implies that \(T_r(k) = T_r(k-1)\). Thus, from the equations provided:

\begin{align}
T_s & \approx A T_s + B P_s, \label{eq:2} \\
(I - A) T_s & \approx B P_s, \label{eq:3} \\
T_s & \approx (I - A)^{-1} B P_s, \label{eq:4} \\
T_s & \approx R P_s. \label{eq:5}
\end{align}

These equations collectively demonstrate how $T_s$ (steady-state temperature) is estimated as a function of power source inputs \(P_s\) using matrices \(A\), \(B\), and \(R\). To derive the $R$ matrix, NMF is utilized to extract $R$ and $P_s$ from steady-state measurements $T_s$ and $P_{T_{s}}(k)$. Our proposed algorithm aims to determine the optimal initialization for NMF to achieve accurate power estimation for each core, $P(k)$.
\section{The Proposed Approach}
\label{Sec:Proposed-Init}
\subsection{BPI Sensitivity to Initialization and Outliers}

NMF, central to BPI, is highly sensitive to initialization and thermal outlier data. NMF's sensitivity arises from solving a non-convex optimization problem with multiple local minima, where initial values for \(R\) and \(P_s\) significantly influence the convergence point, affecting factorization quality and interpretability. Random initialization of these matrices often leads to varying results, necessitating an appropriate initialization method tailored to this specific problem~\cite{math9091006,10.1093/comjnl/bxab103}.

NMF is also sensitive to thermal outliers because outliers distort factorization, leading to poor matrix approximation and degraded feature quality. Outliers increase reconstruction error by fitting anomalous data, reducing factorization accuracy. They slow NMF convergence, requiring more iterations or causing premature convergence to sub-optimal solutions. Additionally, outliers imbalance weight distribution in factorized matrices, resulting in skewed representations of the data's underlying structure~\cite{hafshejani2023initialization}. The NMF initialization is performed once during the offline phase of BPI to determine the SoC matrices $A$ and $B$.

%In the following section, we will address issues with NMF initialization and sensitivity to outliers.

\begin{figure}[htbp]
    \centering
    \vspace{-2mm}
    \includegraphics[width=1.0\linewidth, height=1.0\linewidth]{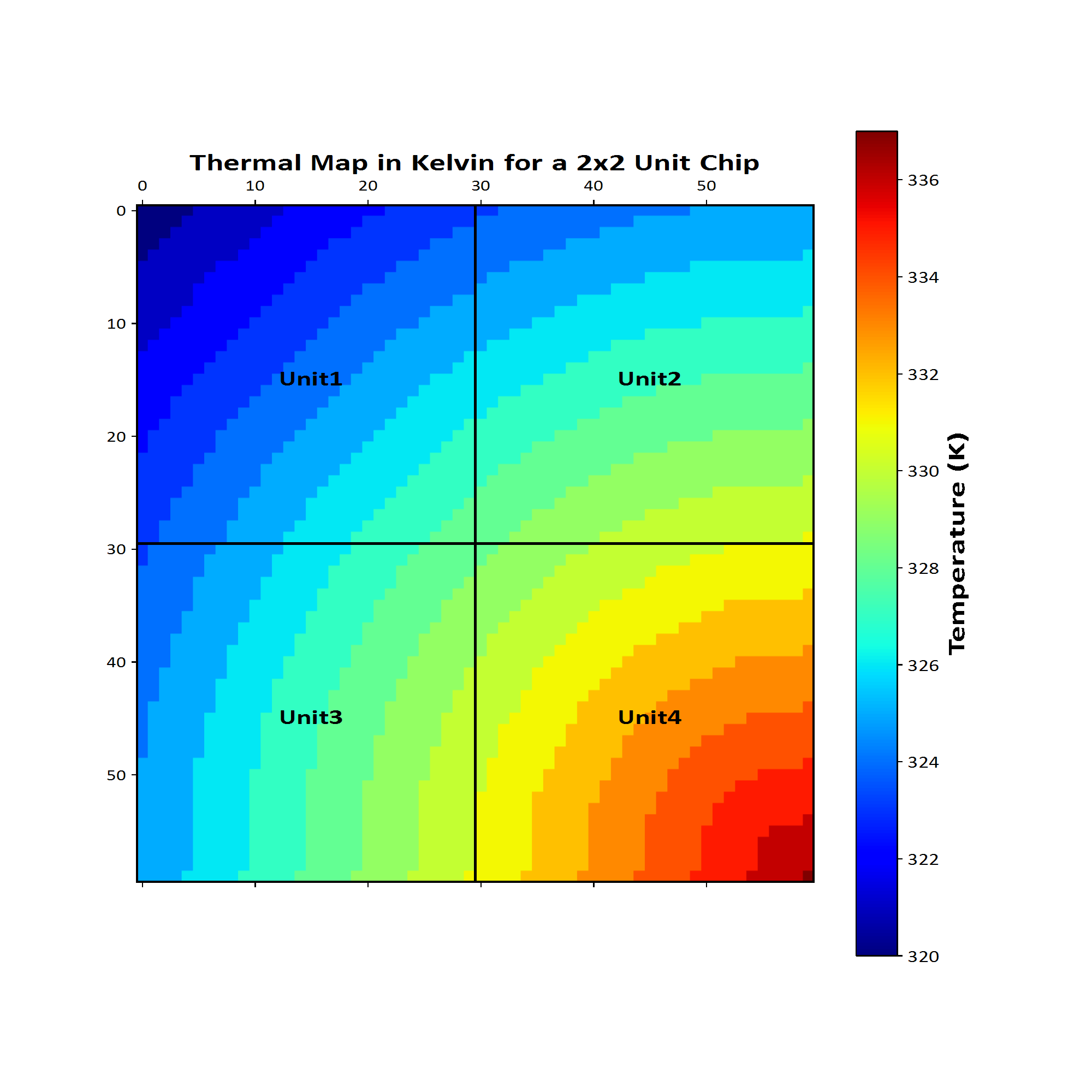}
    \caption{A thermal map for a $2 \times 2$ chip (in Kelvin), where the bottom right unit is activated.}
    \label{fig:2_2_chip}
    \vspace{-2mm}
\end{figure}

%%%%%%%%%%%%%%%%%%%%%%%%%%%%%%%%%%%%%%%%%%%%%%%%%%%%%%%%%%%%%%%%%%%%%%%%%%%%%%%%%%%%%%

\subsection{The Proposed Initialization Approach}
% \label{AA}%Use a better label
% \textcolor{red}{why some equations don't have lable?}
The thermal resistance matrix, $R$, shows how the heat generated in one core affects the temperature of the same and other cores. As illustrated in Figure~\ref{fig:2_2_chip}, a $2 \times 2$ core chip will have a $4 \times 4$ \(R\) matrix, with the element $r_{ij}$ showing the thermal relationship between cores $i$ and $j$. When a core is activated, the highest temperature points cluster around it. 
Our method involves clustering thermal regions and initializing the \(R\) matrix with their centroids, which are the most influential points. From Eqn.~\ref{eq:5}, \( T_s = RP_s \), where \( R \) directly relates to \( T_s \), indicating that any changes in \( P \) will alter \( T_s \). However, in a steady state, where \( T_s \) is fixed, any variation in \( R \) will correspondingly adjust the \( P_s \) matrix and vice versa. %redundant, already said this
%\( R \), representing the thermal transfer or resistance matrix, illustrates how heat from one core influences the temperature of itself and others. 

Given \( R \)'s significant relationship with temperature, initializing the \( R \) matrix with the optimal point from the \( T \) dataset can lead to more accurate convergence of the NMF algorithm towards the true values. Additionally, every point in the original dataset impacts the elements of the \( R \) matrix. 
To effectively manage this, we will employ a robust clustering algorithm, DBSCAN, to categorize the original data into clusters containing the densest points and automatically eliminate outliers. This preprocessing step will aid in the initialization of the \( R \) matrix. We will also set the initial state of \( P \) to match the original steady-state temperature matrix \( T \), allowing \( R \) to adjust \( P \) either upwards or downwards from this baseline to achieve optimal values.

\begin{figure}%[htbp]
\centering
\includegraphics[scale=0.4, trim=0 0 0 0, clip]{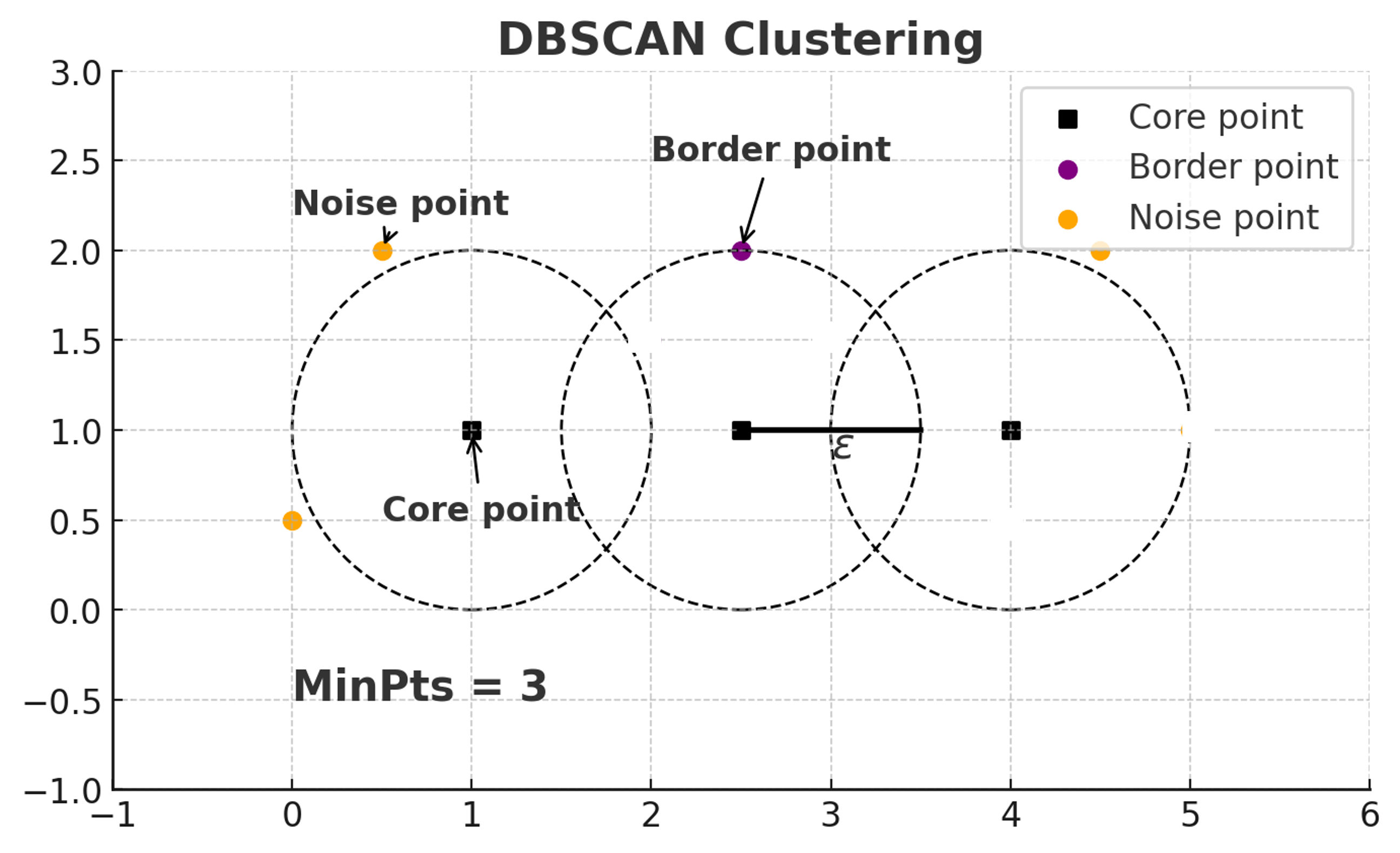}
\caption{DBSCAN Clustering Algorithm Illustration}
\label{fig:DBSCAN_Points}
\end{figure}

DBSCAN is highly efficient in identifying clusters of various shapes and sizes within a dataset, as illustrated in Fig.~\ref{fig:DBSCAN_Points}. It typically uses the Euclidean distance to measure the distance between points:
\begin{equation}
d(p, q) = \sqrt{\sum_{i=1}^{n} (p_i - q_i)^2}
\end{equation}
where \( p \) and \( q \) are two points in an \( n \)-dimensional space. A core point is a point \( p \) with at least \textit{MinPts} points within a distance \( \epsilon \) from it (including \( p \)):
\begin{equation}
|N_\epsilon(p)| \geq \text{MinPts}
\end{equation}
where \( N_\epsilon(p) \) is the set of points within \( \epsilon \) distance from \( p \). A point \( q \) is directly density-reachable from a point \( p \) if \( p \) is a core point and \( q \) is within the distance \( \epsilon \) from \( p \):
\begin{equation}
q \in N_\epsilon(p)
\end{equation}
A point \( q \) is density-reachable from \( p \) if there is a chain of points \( p_1, p_2, \ldots, p_n \) where \( p_1 = p \) and \( p_n = q \), and each point in the chain is directly density reachable from the previous point. 

A core point has at least a minimum specified number of neighboring points (MinPts) within a given radius (\(\epsilon\)), central to a cluster. A border point has fewer neighbors than MinPts within \(\epsilon\) but is near a core point, typically on the cluster's periphery. Noise points do not qualify as either core or border points and are generally considered outliers. DBSCAN segments data based on point density within a specified radius, starting with a random core point and recursively exploring all density-reachable points to expand the cluster. This process continues until all potential clusters are identified. DBSCAN excels in handling noise and discovering clusters with arbitrary shapes, which many clustering algorithms struggle with. However, its effectiveness largely depends on the parameters \(\epsilon\) and MinPts, making proper selection of these parameters crucial for accurate clustering outcomes.

\begin{figure}%[htbp]
\centering
\includegraphics[scale=0.45, trim=0 0 0 0, clip]{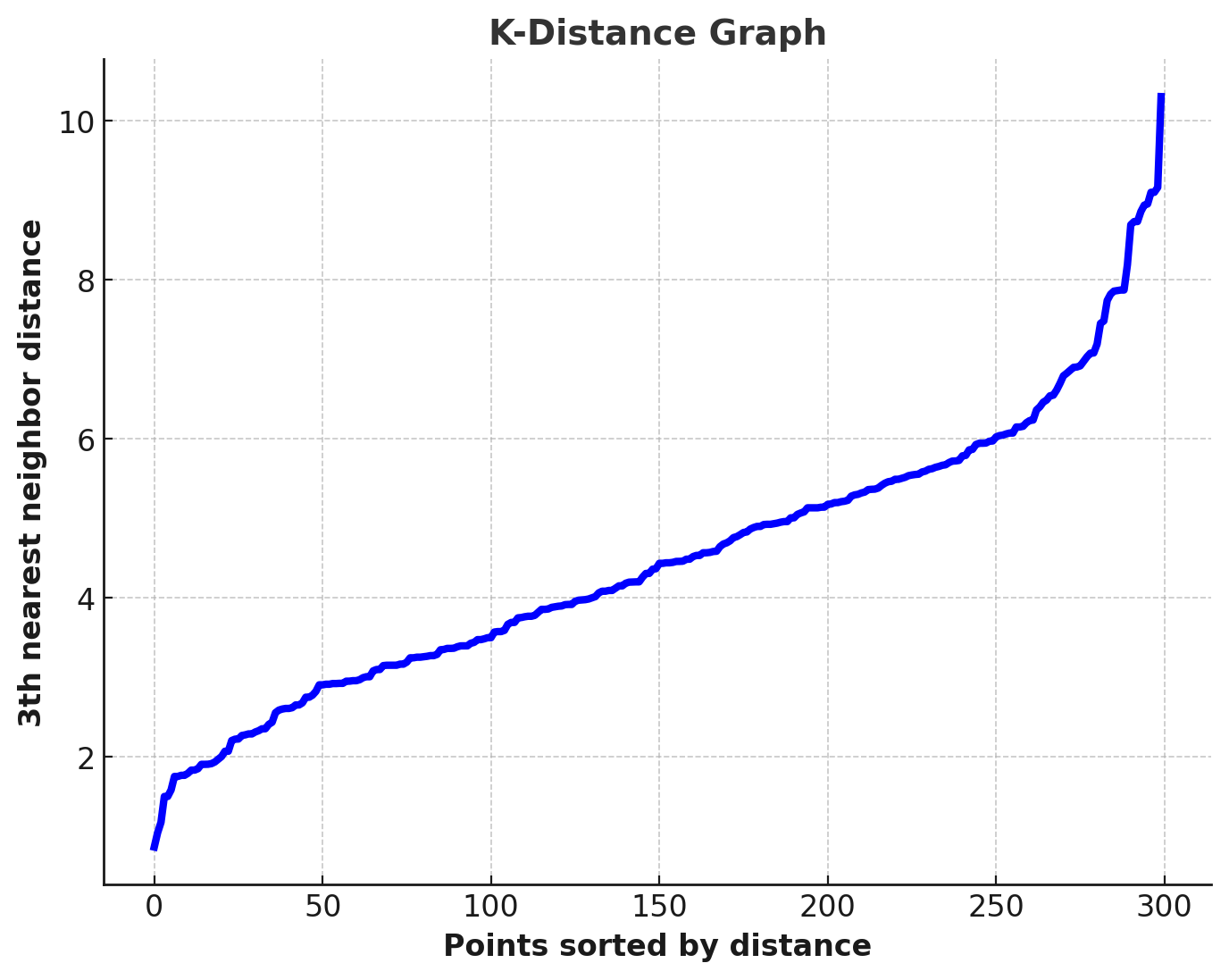}
\caption{K-distance graph to determine the value of \(\epsilon\)}
\label{fig: K-Distance}
\end{figure}

%%%%%%%%%%%%%%%%%%%%%%%%%%%%%%%%%%%%%%%%%%%%%%%%%%%%%%%%%%%%%%
To estimate the optimal \(\epsilon\) for DBSCAN, the k-distance graph is used, which plots the distance to the k-th nearest neighbor for each data point in descending order. The "elbow point" on this graph, where there is a significant change in slope, indicates the optimal \(\epsilon\), as shown in Fig.~\ref{fig: K-Distance}. This point signifies a natural threshold between dense clusters and isolated points. The detailed structure of the proposed approach is provided in Algorithm~\cref{alg:DBSCAN_BPI}.

\begin{algorithm}[H]
\caption{\textit{Proposed Approach}: Initialization with DBSCAN Clustering}
\label{alg:DBSCAN_BPI}
\begin{algorithmic}[1]
    \STATE \textbf{Input:} Steady-state temperature data $T_s$, Number of cores $n$, Total power at each interval $P_{T_s}(k)$
    \STATE Determine the optimal \(\epsilon\) using the k-distance graph.
    \STATE Set \textit{MinPts} to at least \(D + 1\), where \(D\) is the number of features.
    \STATE Apply DBSCAN to $T_s$ to identify clusters and remove outliers.
    \STATE Initialize \(R\) with the centroids of the clusters.
    \STATE Initialize \(P_s\) using $T_s$ after outlier removal.
    \STATE Perform NMF with the proposed initialization.
    \STATE \textbf{Output:} System matrices \(A\), \(B\), and \(R\).
\end{algorithmic}
\end{algorithm}

%%%%%%%%%%%%%%%%%%%%%%%%%%%%%%%%%%%%%%%%%%%%%%%%%%%%%%%%%%%%%%%%%%%%%%%%%%%%%%%%%%%%%%%
\section{Experiments and Results }
\label{sec:results}
\subsection{Experimental setup}

The effectiveness of the proposed approach is validated using the HotSpot thermal simulator~\cite{b11,b12} for four different floor plans described in Table~\ref{table:floorplans}. 

\begin{table}[h]
\centering
\caption{The floorplans with unit counts, architecture types, and power budgets}
\label{table:floorplans}
\begin{tabular}{| >{\centering\arraybackslash}p{2.9cm} | >{\centering\arraybackslash}p{0.6cm} | >{\centering\arraybackslash}p{1.7cm} | >{\centering\arraybackslash}p{1.8cm} |}
\hline
\textbf{Floorplans} & \textbf{Units} & \textbf{Architecture} & \textbf{Power Budget} \\ \hline
$2 \times 2$ mesh (FP1) & 4 & Homogeneous & 60 W \\ \hline
$2 \times 4$ mesh (FP2) & 8 & Homogeneous & 60 W \\ \hline
$4 \times 4$ mesh (FP3) & 16 & Homogeneous & 60 W \\ \hline
Big LITTLE+GPU (FP4) & 6 & Heterogeneous & 10 W \\ \hline
\end{tabular}
\end{table}

% left bot right top

\begin{figure}%[htbp]
\centering
\includegraphics[scale=0.17, trim=0 0 0 0, clip]{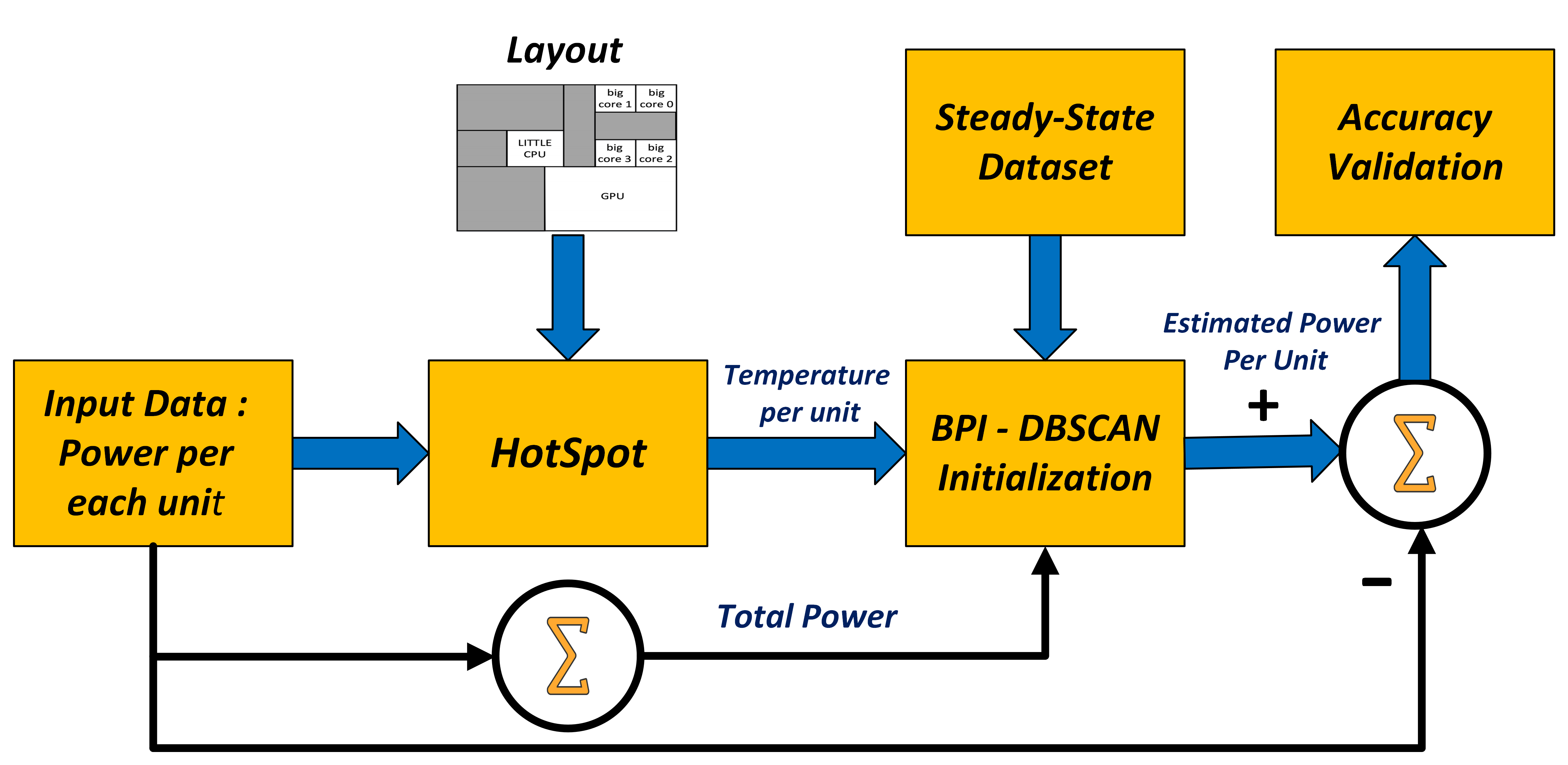}
\caption{The verification and testing flow of the proposed approach}
\label{fig:HotSpot}
\end{figure}

% \textcolor{red}{figure 5 is too big and figures 9-10 too small. If more space is needed, the heights of figures 7 and 8 can be halved!}

As depicted in Fig.~\ref{fig:HotSpot}, the HotSpot simulator receives per-unit power traces from a given design layout and generates the corresponding per-unit temperature traces. These temperature outputs are then utilized as inputs by the proposed approach alongside total power data to calculate the estimated per-unit power. Finally, the accuracy of the proposed approach is determined by comparing the initial per-unit power traces, which were inputs to the HotSpot simulator, with the power estimates generated by the proposed approach.

\begin{figure}%[htbp]
\centering
\includegraphics[width=0.65\linewidth]{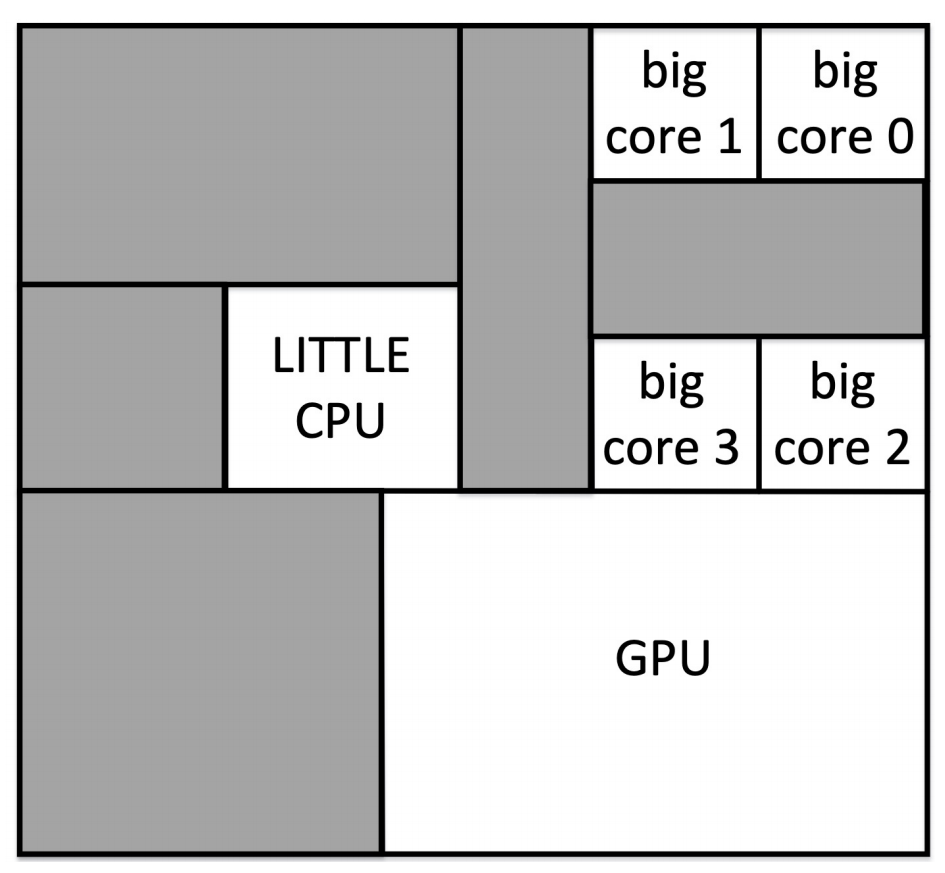}
\caption{The layout of the big.LITTLE+GPU SoC, utilized for testing the proposed approach~\cite{b14}}
\label{fig: Hetro_Arch}
\end{figure}

The proposed approach is compared with the original BPI approach, which was the first to introduce BPI for fine-grain power estimation. BPI relies on NMF to identify the \( B \) matrix. However, the accuracy of the NMF output is sensitive to the initialization step, which in this case was performed based on the identity matrix as an initial \( R \) matrix. Therefore, it is also compared against another version of BPI~\cite{b6} that improved the initialization step by utilizing steady-state temperatures, which is referred to as BPISS. Furthermore, the proposed approach is validated on a heterogeneous architecture to demonstrate that the proposed approach is robust in terms of both accuracy and computation time when transitioning from a homogeneous to a heterogeneous architecture that was depicted in Fig.~\ref{fig: Hetro_Arch}.
%%%%%%%%%%%%%%%%%%%%%%%%%%%%%%%%%%%%%%%%%%%%%%%%%%%%%%%%%%%%%%%%%%%%%%%%%%%%%%%%%%%%%%%%

\subsection{HotSpot Simulation Results}
\begin{figure}%[htbp]
\centering
\vspace{-5mm}
\includegraphics[width=1.02\linewidth, trim=0 0 0 37, clip]{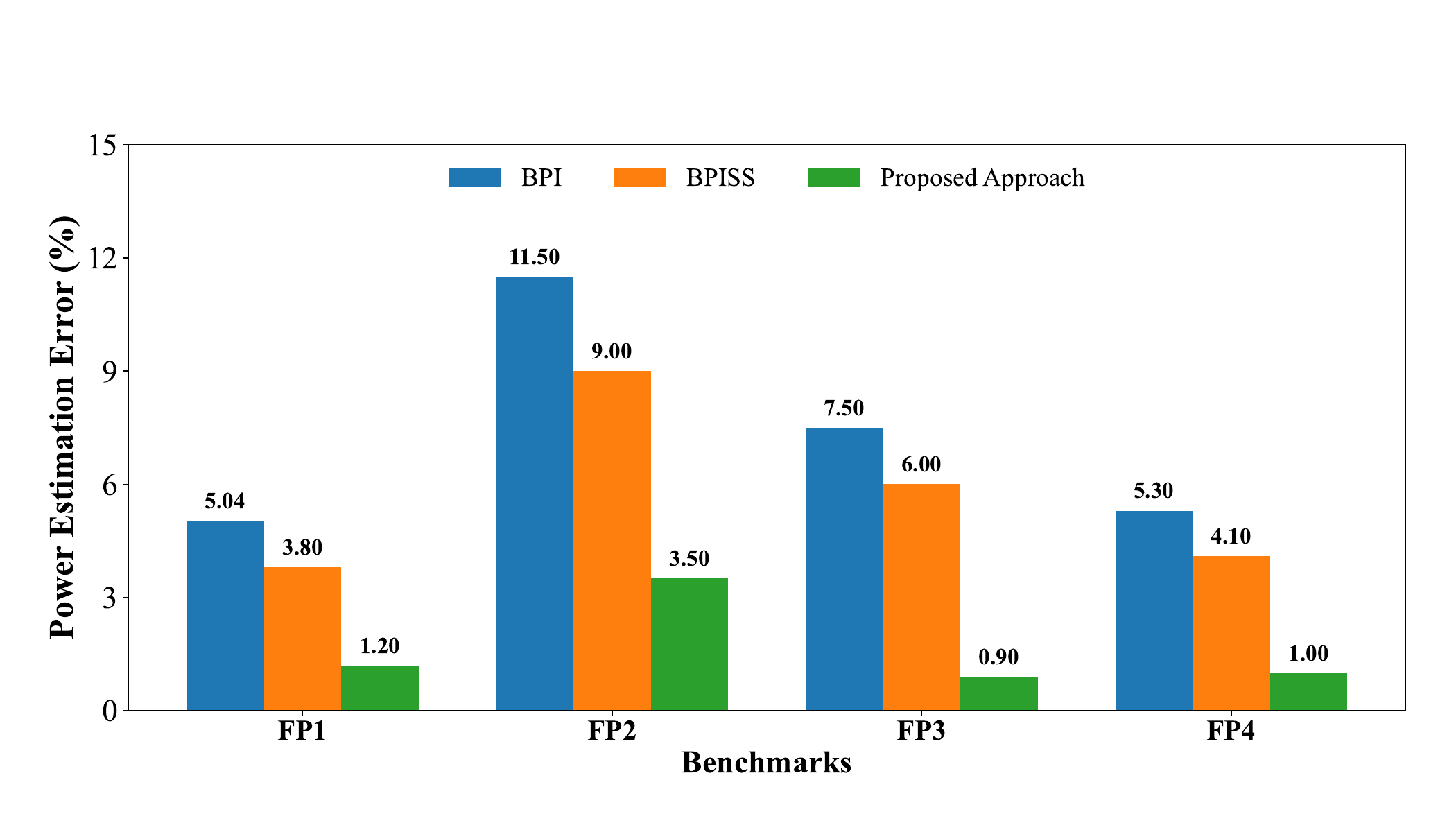} 
\caption{Comparison of Power Estimation Errors Across Different Algorithms and Benchmarks} %This figure illustrates the power estimation error percentages for different benchmark configurations (2x2 mesh, 2x4 mesh, 4x4 mesh, and big.LITTLE+GPU) using three different algorithms: BPI, BPISS, and DBSCAN-BPI. The results highlight the varying performance of each algorithm in estimating power consumption accurately across different system configurations.}
\label{fig:power_error}
\vspace{-2mm}
\end{figure}

Figure~\ref{fig:power_error} presents a comparative analysis of power estimation errors associated with two different approaches mentioned in the related work: BPI and BPISS, and the proposed approach. The analysis is carried out on four different benchmarks, detailed in Table~\ref{table:floorplans}.

\textbf{Prediction Accuracy:} The proposed method consistently outperforms both BPI and BPISS in terms of power estimation accuracy across all tested benchmarks. In the case of FP1, the proposed approach brings the error down to around $1.2\%$, whereas BPI and BPISS exhibit higher errors of about $5\%$ and $3.8\%$, respectively. For FP2, the error is reduced to approximately $3.5\%$ with the proposed approach, which is notably better than the $11.5\%$ error for BPI and $9\%$ for BPISS. Similarly, for FP3, the error drops to around $0.9\%$, in contrast to the $7.5\%$ and $6\%$ errors seen with BPI and BPISS, respectively. Lastly, for FP4, our proposed method records an error of roughly $1\%$, outperforming BPI's ($5.3\%$) and BPISS's ($4.1\%$).

The improvement is particularly noticeable as the size of the benchmark increases, indicating the scalability of the proposed approach. The algorithm maintains a higher accuracy estimation rate when transitioning from homogeneous to heterogeneous processors, demonstrating its robustness and reliability in handling complex and larger architectures.

\begin{figure}%[htbp]
\centering
\vspace{-3mm}
\includegraphics[width=\linewidth]{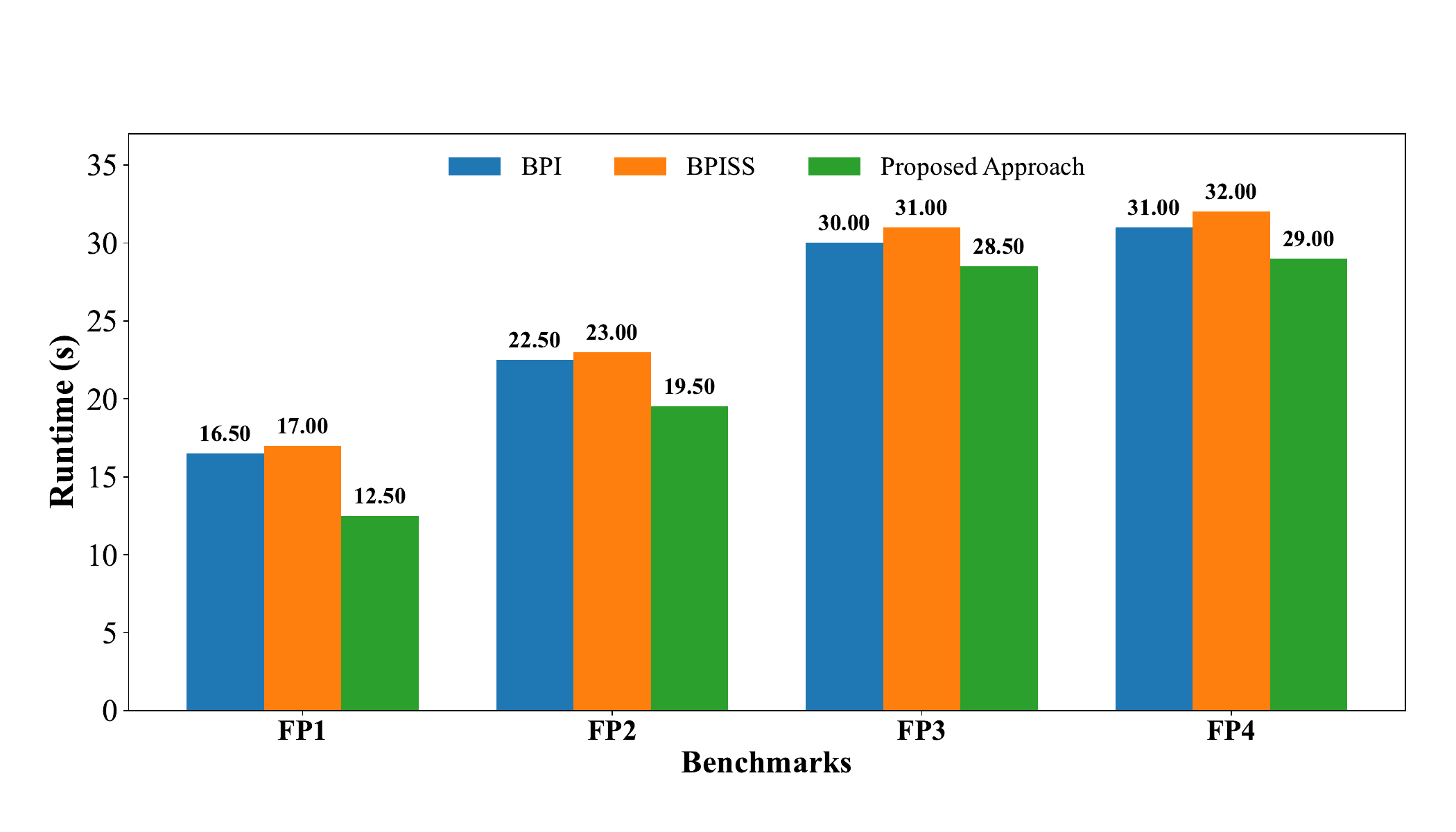}
\caption{Runtime Comparison of Different Algorithms Across Various Benchmarks}
\label{fig: Rune_time}
\vspace{-2mm}
\end{figure}

Figure~\ref{fig: Rune_time} presents a comparative analysis of the runtime for the three algorithms: BPI, BPISS, and the proposed approach across the four benchmarks.

\textbf{Runtime Analysis:} The proposed approach generally shows a lower runtime compared to BPISS and is comparable to BPI in most benchmarks and helps reduce the overall computational complexity.

\textbf{Improvement over Traditional Methods:} Traditional BPI methods exhibit slightly lower run times but suffer from higher estimation errors, indicating a trade-off between speed and accuracy. BPISS shows the highest runtime, increasing computational overhead. 

\textbf{Scalability and Performance:} Our proposed approach scales well with increasing benchmark sizes, maintaining efficient runtimes while handling larger and more complex architectures. This scalability ensures that the algorithm can be applied to more complex systems without a significant increase in computational resources or time.

% add the simulation plot figures and discussion for it 

\subsection{Comparison of Actual Power with Estimated Power}
\begin{figure}%[h]
\centering
\includegraphics[scale=0.7, trim=0 0 0 0, clip]{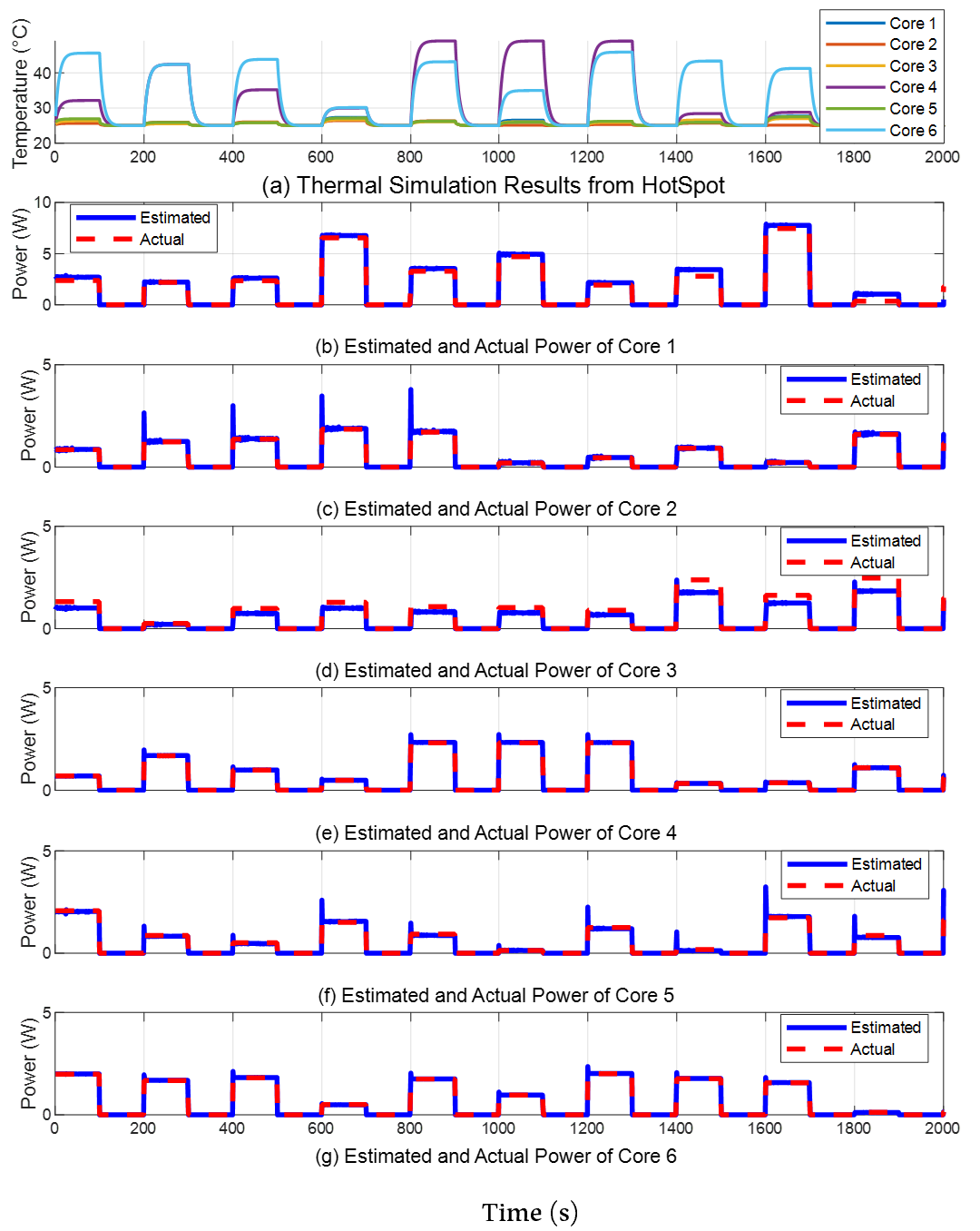}
\caption{The proposed BPI algorithm was validated using the HotSpot simulator based on the heterogeneous floorplan}
\label{fig:Hetro23}
\end{figure}

As depicted in Figure~\ref{fig:Hetro23}(a), illustrates the thermal measurements for each of the cores over time for heterogeneous processors (FP4). The temperatures fluctuate as the cores experience varying states of stress and periods of idleness. These temperature variations are crucial for understanding the thermal dynamics and behavior of the processors under different operational conditions.

Figures~\ref{fig:Hetro23} provide a detailed comparison between the power estimates derived from our proposed approach and the actual power inputed to each core as simulated by HotSpot. This comparison is essential to validate the accuracy and reliability of the proposed approach in estimating power consumption. The results demonstrate that the proposed approach closely matches the estimated and actual power values. This close correlation underscores the effectiveness of our approach in providing accurate power estimates, which is vital for optimizing power management and improving the overall efficiency of the SoCs.

% \begin{figure}%[h]
%     \centering
%     \includegraphics[scale = 0.65]{Images/Simualtion_Fig1.png}
%     \caption{The proposed BPI algorithm was validated using the HotSpot simulator using the 4-core floorplan.}
%     \label{fig:4_Core}
% \end{figure}
% left bot right top

% \begin{figure}[h!]
%   \centering
%   \begin{subfigure}{0.5\textwidth}
%     \includegraphics[width=\linewidth]{Images/Simualtion_Fig1.png}
%     \caption{First subfigure}
%   \end{subfigure}
%   \hfill
%   \begin{subfigure}{0.5\textwidth}
%     \includegraphics[width=\linewidth]{Images/Simualtion_Fig3.png}
%     \caption{Second subfigure}
%   \end{subfigure}
%   \caption{Main caption}
%   \label{fig:yourlabel}
% \end{figure}
\section{Conclusions}
\label{sec:conl}

Our approach markedly improves the accuracy and robustness of the Blind Power Identification (BPI) technique by employing DBSCAN for the optimal initialization of Nonnegative Matrix Factorization (NMF), thereby mitigating its sensitivity to variations in datasets and outliers. Our comprehensive simulations have shown that the proposed approach provides superior accuracy in fine-grain power estimation across different multicore processor floorplans, including both homogeneous and heterogeneous architectures. Specifically, error rates in a four core processor were reduced by $67\%$ compared to conventional BPI techniques and by $24\%$ compared to the BPISS method. Future work could focus on integrating the proposed approach with thermal management and power control techniques, such as dynamic voltage and frequency scaling (DVFS) and thermal throttling, to create a comprehensive solution for managing power and thermal behavior in SoCs. Additionally, our approach can be combined with security techniques to identify malicious attacks based on thermal behavior.

\section*{Acknowledgments} 
%The authors thank Dr. Mostafa Abdelrahim of Qualcomm Inc.\ for providing codes from his BIC paper (ISQED'2022)~\cite{9806243}. 
The authors would like to thank the anonymous reviewers. 
This work has been partially funded by NSF grants 2219679 and 2219680.
%%%%%%%%%%%%%%%%%%%%%%%%%%%%%%%%%%%%%%%%%%%%%%%%%%%%%%%%%%%%%%%%%%%%%%%%%%%%%%%%%%%%%

% \bibliographystyle{IEEEtran} % IEEE bibliography style
\bibliography{References} % Use the single .bib file

\end{document}